%% file: MSBv2.tex
\documentclass[aps,prl,showpacs,twocolumn,superscriptaddress]{revtex4}
\usepackage{multirow}
\usepackage{graphicx}
\usepackage{amsmath}
\usepackage{pslatex}

\newcommand{\sqrtsNN}   {\mbox{$\sqrt{\mathrm{s}_{_{\mathrm{NN}}}}$}}

\newcommand{\xim}   {$\Xi^-$}

\newcommand{\lam}   {$\Lambda$}
\newcommand{\ks}    {$\mathrm{K}^{0}_{S}$}
\newcommand{\omm}   {$\Omega^-$}

\newcommand{\pt}    {$p_T$}
\def \auau          {$\mathrm{Au}+\mathrm{Au}$ }

\def \xxi           {$\Xi^{-}+\overline{\Xi}^{+}$ }
\def \oom           {$\Omega^{-}+\overline{\Omega}^{+}$ }
\def \lla           {$\Lambda+\overline{\Lambda}$}
\begin{document}

\title{Multi-strange baryon elliptic flow in \auau\ collisions at
\sqrtsNN\ =200 GeV}

\input{sci-apr05.tex}


\begin{abstract}
We report on the first measurement of elliptic flow $v_2$(\pt) of
multi-strange baryons \xxi and \oom in heavy-ion collisions. In
minimum bias \auau\ collisions at \sqrtsNN = 200 GeV, a significant
amount of elliptic flow, comparable to other non-strange baryons, is
observed for multi-strange baryons which are expected to be
particularly sensitive to the dynamics of the partonic stage of
heavy-ion collisions. The \pt\ dependence of $v_2$ of the
multi-strange baryons confirms the number of constituent quark
scaling previously observed for lighter hadrons. These results
support the idea that a substantial fraction of the observed
collective motion is developed at the early partonic stage in
ultra-relativistic nuclear collisions at RHIC.
\end{abstract}


\maketitle


Lattice QCD calculations, at vanishing or finite net-baryon density,
predict a transition from the deconfined thermalized partonic matter
Quark Gluon Plasma (QGP) to ordinary hadronic matter at a critical
temperature $T_c \approx 150 - 180$ MeV \cite{karsch,fodor02}.
Measurements of hadron yields in the intermediate ($2\lesssim p_{T}
\lesssim6$ GeV/$c$) and high ($p_{T}\gtrsim6-8$ GeV/$c$) transverse
momentum $p_{T}$ region indicate that dense matter has been produced
in \auau\ collisions at RHIC
\cite{star_1s,star_2s,star_2dau,phenix_1s,phenix_2s2,phenix_2dau,phobos_2dau,brahms_2dau}.
Furthermore, previous measurements of elliptic flow of hadrons
indicate that the matter created at RHIC is also strongly
interacting~\cite{StarPidV2,StarK0sLamV2}. Thus, in the early stage
of the collision, dense and strongly interacting matter will lead to
collective effects among constituents such as transverse collective
motion. If these interactions occur frequently enough, the system
will finally reach thermalization. Due to the initial spatial
anisotropy of the system in non-central collisions, an elliptic
component of the collective transverse motion should also be
present. Collectivity is cumulative throughout the whole collision
and should survive the hadronization process~\cite{Kolb03,Teaney01};
therefore, the amount of transverse flow observed in the final state
will have a contribution from the pre-hadronic, i.e. partonic,
stage.

Early dynamic information might be masked by later hadronic
rescatterings. Multi-strange baryons with their large mass and
presumably small hadronic cross
sections~\cite{vanHecke98,Bass99,Cheng03,Biagi81,Muller72}, should
be less sensitive to hadronic rescattering in the the later stages
of the collision and therefore a good probe of the early stage of
the collision~\cite{StarMSB130}. Indeed, a systematic study of
hadron \pt\ spectra from high-energy heavy ion collisions, using a
hydrodynamically inspired model, shows that multi-strange baryons
thermally freeze-out close to the point where chemical freeze-out
occurs with $T_{\rm ch} \sim 160$ MeV~\cite{StarMSB130,na49om} which
at these collision energies coincides with the critical temperature
$T_c$~\cite{karsch,fodor02}. This may mean that multi-strange
baryons are not, or much less, affected by hadronic rescatterings
during the later stage of heavy ion
collisions~\cite{vanHecke98,Bass99}. Their observed transverse flow
would then primarily reflect the partonic flow. Moreover, elliptic
flow is in itself considered to be a good tool for understanding the
properties of the early stage of the
collisions~\cite{sorge99,Ollitrault92}, primarily due to its
self--quenching nature. Elliptic flow is generated from the initial
spatial anisotropy of the system created in non-central collisions
by rescatterings among the constituents of the system. The generated
elliptic flow will reduce the spatial anisotropy of the system and
quench its own origin. Thus multi--strange baryon elliptic flow
could be a valuable probe of the initial partonic system.

In this Letter we present the first results on elliptic flow of
multi-strange baryons \xxi and \oom from \auau\ collisions at
\sqrtsNN = 200 GeV, as measured with the STAR
detector~\cite{StarNIM}. About 2 million events from \auau\
collisions collected with a minimum bias trigger are used in this
analysis. Multi-strange baryons are reconstructed via their decay
topology: $\Xi \rightarrow \Lambda + \pi$ and $\Omega \rightarrow
\Lambda + K$ with the subsequent decay of $\Lambda \rightarrow p +
\pi$ as described in~\cite{StarMSB130}. Charged tracks were
reconstructed in the STAR Time Projection Chamber
(TPC)~\cite{TpcNIM}. Simple cuts on geometry, kinematics and
particle identification via specific ionization are applied to
reduce the combinatorial background. A detailed description of the
analysis procedure can be found in \cite{Huilambda,StarMSB130}.

\begin{figure} [h]
\includegraphics[width=0.48\textwidth]{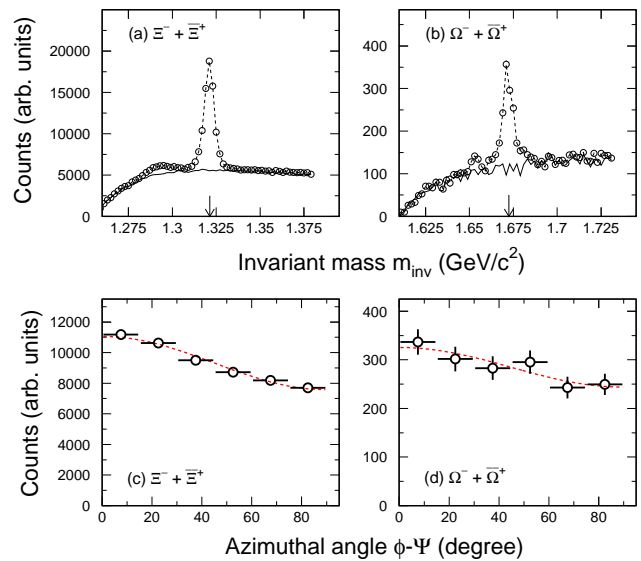}
\caption{(color online). (a) \xxi and (b) \oom invariant mass
distribution from minimum bias (0--80\%) \auau\ collisions at
\sqrtsNN = 200 GeV. The solid line shows the combinatorial
background as estimated from a same event rotating method (see text
for details). Azimuthal distributions with respect to the event
plane of the (c) \xxi and (d) \oom raw yields. Dashed lines
represent the fit results. All plots shown include \xxi and \oom in
the transverse momentum range $1 < p_T < 4$~GeV/$c$.}
\label{Fig1:InvMassXiOm}
\end{figure}


Figure~\ref{Fig1:InvMassXiOm} shows the invariant mass distribution
for (a) \xxi and (b) \oom candidates from minimum bias collisions
(0--80\% of the total hadronic cross-section). The \xxi and \oom
signals appear as clear peaks around the rest masses (indicated by
the vertical arrows) in the invariant mass distribution, above a
combinatorial background. The combinatorial background of
uncorrelated decay candidates under the peak can be determined by
sampling the regions on both sides of the peak. It can also be
reproduced by rotating the $\Lambda$ candidates by $180^{\circ}$ in
the transverse plane and then reconstructing the $\Xi$ and $\Omega$
candidates. The rotation of the $\Lambda$ breaks the correlation in
the invariant mass and therefore mimics the background of
uncorrelated decay pairs. Both background determination methods
provide consistent results. In Fig.~\ref{Fig1:InvMassXiOm} (a) and
(b), the combinatorial background as calculated from the rotation
method is shown as solid lines. Outside the region of the
corresponding mass peak, the rotation method describes the
background well. The residual {\it bump} at lower invariant mass
than the peak in Fig.~\ref{Fig1:InvMassXiOm} (a) can be understood
as fake $\Xi$ candidates being reconstructed as $\Xi_{\rm
fake}$($\pi_{\Lambda}$,$\Lambda_{\rm fake}$($\pi_{\rm
random}$,$p_{\Lambda}$)), where $\pi_{\Lambda}$ and $p_{\Lambda}$
are the daughters of a real $\Lambda$ and $\pi_{\rm random}$ is a
random $\pi$. The real correlation between $\pi_{\Lambda}$ and
$p_{\Lambda}$ remains in the $\Xi_{\rm fake}$ reconstruction
resulting in the observed bump in the $\Xi$ invariant mass
distribution. A similar mis-association happens in the $\Omega$ case
with the addition of the $\pi_{\Lambda}$ being misidentified as a
kaon. Our studies have shown that this residual correlation does not
affect the signal peak. The raw yields are then extracted from the
invariant mass distribution by counting the number of entries in the
mass peak above the estimated background.

The elliptic flow $v_{2}$ is calculated from the distribution of
particle raw yields as a function of azimuthal angle $\phi$ with
respect to the event plane angle $\Psi$. The $\Xi$ and $\Omega$
candidates are divided in $\phi-\Psi$ bins, and the raw yields for
each bin are extracted from the invariant mass distributions as
described above. The event plane angle $\Psi$ is used as an estimate
of the reaction plane angle~\cite{Starv2130,RPresolution}. Here, the
event plane is determined from the azimuthal distribution of charged
primary tracks with $0.2<p_T<2.0$ GeV/$c$ and pseudo-rapidity
$|\eta|<1.0$. To avoid autocorrelations, tracks associated with a
$\Xi$ or $\Omega$ candidate are explicitly excluded from the event
plane calculation. Figure~\ref{Fig1:InvMassXiOm} shows the azimuthal
distributions of raw yields for (c) \xxi and (d) \oom with respect
to the event plane from the minimum bias collisions in the $1<p_T<4$
GeV/$c$ range. To reduce the statistical uncertainties in the $\Xi$
and $\Omega$ signal extraction and because of the
$\cos~2(\phi-\Psi)$ dependence of $v_{2}$, we have folded around
$\pi/2$ the candidates in the $\pi/2<\phi-\Psi<\pi$ range into the
$\pi/2>\phi-\Psi>0$ range. The distributions exhibit a clear
oscillation with azimuthal angle $\phi-\Psi$ for both $\Xi$ and
$\Omega$ particles indicating the presence of significant elliptic
flow. Dashed lines are results from fitting a function
$\frac{dN}{d(\phi-\Psi)}=A[1+2v_2~\cos~2(\phi-\Psi)],$ where $A$ is
the normalization constant. Furthermore, we note that the amplitude
of the oscillation for the $\Xi$ and $\Omega$ are of similar
magnitude indicating that their $v_2$ is similar, as will be
discussed later. The finite resolution in the event plane
determination smears out the azimuthal distributions and leads to a
lower signal in the apparent anisotropy~\cite{RPresolution}. We
determine the event plane resolution by dividing each event into
random sub-events and determine the correction factor to be $1/0.72$
for minimum-bias collisions. In the following, all numbers reported
on $v_2$ are corrected for this resolution. Systematic uncertainties
in $v_2$ were studied by comparing the background determination
methods described above and by changing the cuts used in the $\Xi$
and $\Omega$ reconstruction. For the $\Xi$, the estimated absolute
systematic uncertainties are 0.02 for the lowest \pt\ bin and
smaller than 0.01 for all other \pt\ bins. For the $\Omega$, the
absolute systematic uncertainty is 0.04 for both measured transverse
momentum bins. Correlations unrelated to the reaction plane
(non--flow effects) can modify the apparent
$v_2$~\cite{StarK0sLamV2}. Non-flow contributions for multi-strange
baryons were not studied yet, but are expected to be similar to
those calculated for $\Lambda$ ($\sim -0.01$ at \pt=1 GeV/$c$ and
$\sim -0.04$ at \pt=2.5 and 4.0 GeV/$c$)~\cite{StarK0sLamV2}.

\begin{figure}[tb]
\includegraphics[width=0.45\textwidth]{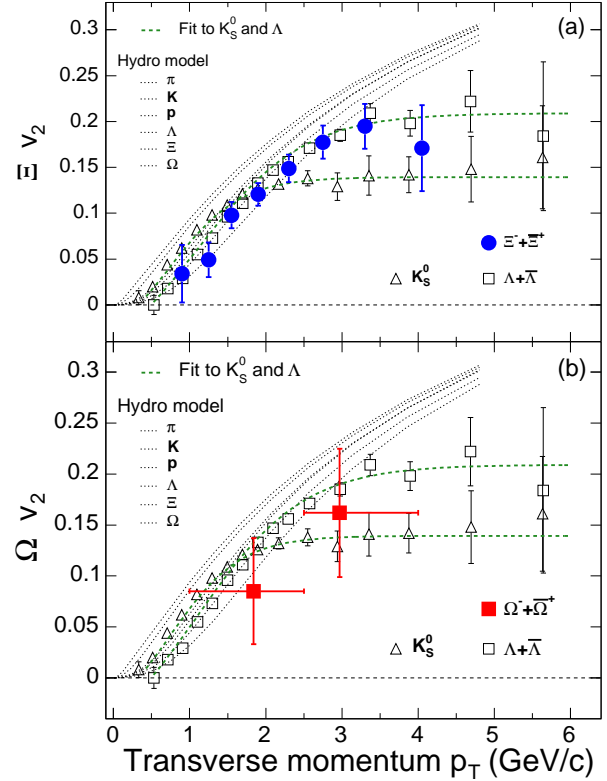}
\caption{(color online). $v_2$(\pt) of (a) \xxi and (b) \oom from
200 GeV \auau\ minimum bias collisions.  The $v_2$ of \ks\ and \lam\
\cite{StarK0sLamV2} are also shown as open symbols and the results
of the fits \cite{xin04} are shown as dashed-dot lines. Hydrodynamic
model calculations \cite{Pasi01} are shown as dotted curves for
$\pi$, $K$, $p$, \lam, \xim, and \omm masses, from top to bottom,
respectively.} \label{Fig2:V2PtXiOm}
\end{figure}

Figure~\ref{Fig2:V2PtXiOm} shows the results of the elliptic flow
parameter $v_2(p_T)$ for multi-strange baryons (a) \xxi\ and (b)
\oom\ from minimum bias (0--80\%) \auau\ collisions. As a reference,
the open symbols represent the published~\cite{StarK0sLamV2} \ks\
and \lam\ $v_2(p_t)$ from the same event class. As guideline,
results of the fit~\cite{xin04} to $v_2(p_T)$ of \ks\ and \lam\ are
shown as dot-dashed-lines. Hydrodynamic model calculations using an
Equation Of State (EOS) with a phase transition at $T_{\rm c}=165$
MeV and a thermal freeze-out at $T_{\rm fo} \sim 130$
MeV~\cite{Pasi01} are shown as dotted lines, from top to bottom, for
$\pi$, $K$, $p$, $\Lambda$, $\Xi$, and $\Omega$, respectively. The
expected mass ordering in hydrodynamics of $v_2(p_T)$ is observed
with lighter particles having larger $v_2(p_T)$ than heavier
particles.

First, we observe in Fig.~\ref{Fig2:V2PtXiOm} (a) that for $\Xi$,
the $v_2$ increases with $p_T$ reaching a saturation value of $\sim
18\%$ at $p_T\sim3.0$ GeV/$c$. This is similar to the result for
$\Lambda$ baryons~\cite{StarK0sLamV2}. In the lower \pt\ region
($p_{T}<2.5$ GeV/$c$), the $\Xi$ results are in agreement with the
hydrodynamic model prediction~\cite{Pasi01}. Second, we observe in
Fig.~\ref{Fig2:V2PtXiOm} (b) that the values of $v_2$ for the
$\Omega$, are clearly non-vanishing although they have larger
statistical uncertainties due to their smaller abundance. Over the
measured $p_T$ range and considering the statistical uncertainties,
the $v_2$ of the $\Omega$ is non-zero with 99.73\% confidence level
($3 \sigma$ effect). The $\Omega$ $v_{2}$ values are, within
uncertainties, consistent with those measured for the $\Xi$,
indicating that even the triply-strange baryon $\Omega$ has
developed significant elliptic flow in \auau\ collisions at RHIC. In
the scenario where multi-strange baryons are less affected by the
hadronic stage~\cite{StarMSB130} and where $v_2$ develops primarily
at the early stage of the collision ~\cite{sorge99,Ollitrault92},
the large $v_2$ of multi-strange baryons reported in this paper
shows that partonic collectivity is generated at RHIC.

Previously, a particle type (baryon versus meson) difference in
$v_2(p_t)$ was observed for $\pi$ and $p$~\cite{PhenixV2} as well as
for \ks\ and \lam~\cite{StarK0sLamV2} at the intermediate \pt\
region. The present results on the $\Xi$ $v_2(p_t)$ follow closely
the ones for $\Lambda$ confirming that this observed particle type
difference is a meson-baryon effect rather than a mass effect. This
particle type dependence of the $v_2(p_T)$ is naturally accounted
for by quark coalescence or recombination
models~\cite{Molnar03,Fries03,coal}. In these hadronization models,
hadrons are formed dominantly by coalescing massive quarks from a
partonic system with the underlying assumption of collectivity among
these quarks. Should there be no difference in collectivity among
$u$-, $d$-, and $s$-quarks near hadronization, these models predict
a universal scaling of $v_2$ and the hadron transverse momentum
$p_T$ with the number of constituent quarks ($n_{q}$). This scaling
has previously been observed to hold within experimental
uncertainties for the \ks\ and the \lam\ when $p_T/n_{q}\ge 0.7$
GeV/$c$~\cite{StarK0sLamV2}.

\begin{figure}[tb]
\includegraphics[width=0.48\textwidth]{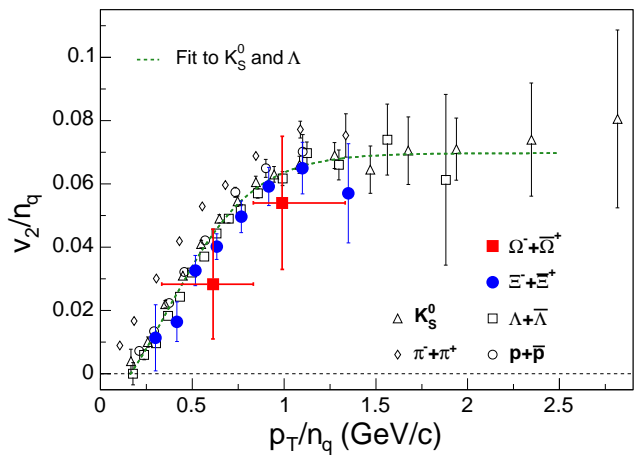}
\caption{(color online). Number of quark ($n_q$) scaled $v_2$ as a
function of scaled $p_T$ for \xxi (filled circles) and \oom (filled
squares). Same distributions also shown for $\pi^+ + \pi^-$ (open
diamonds), $p+\bar{p}$ (open triangles) \cite{PhenixV2}, \ks\ (open
circles), \lla\ (open squares) \cite{StarK0sLamV2}. All data are
from 200 GeV \auau \ minimum bias collisions. The dot-dashed-line is
the scaled result of the fit to \ks\ and $\Lambda$ \cite{xin04}.}
\label{Fig3:v2nq}
\end{figure}

The $n_{q}$-scaled $v_2$ versus the $n_{q}$-scaled \pt\ are shown in
Fig.~\ref{Fig3:v2nq} for $\pi^{-}+\pi^{+}$ (open diamonds),
$p+\bar{p}$ (open circles)~\cite{PhenixV2}, $K^0_S$ (open
triangles), \lla (open squares)~\cite{StarK0sLamV2}, \xxi (filled
circles) and \oom (filled squares). Except for pions, all hadrons
including $\Xi$ and $\Omega$ scale well within statistics. The
discrepancy in the pion $v_2$ may in part be attributed to its
Goldstone boson nature (its mass is smaller than the sum of its
constituent quark masses) or to the effects of resonance decays (a
large fraction of the measured pions will come from the decays of
resonances at higher \pt)~\cite{xin04,ko04}. This further success of
the coalescence models in describing the multi-strange baryon
$v_2(p_T)$ also lends strong support to the finding that
collectivity developed in the partonic stage at RHIC. In addition,
the good agreement of $v_{2}(p_{T}/n_{q})/n_{q}$ for $p(uud)$,
$\Lambda(uds)$, $\Xi(dss)$ and $\Omega(sss)$ further supports the
idea that the partonic flow of $s$ quarks is similar to that of $u,
d$ quarks. Future measurements with higher statistics, specially for
the $\Omega$, will allow for a more quantitative comparison.


In summary, we reported the STAR results on multi-strange baryon,
\xxi and \oom, elliptic flow $v_2$ from minimum bias \auau\
collisions at \sqrtsNN = 200 GeV. The observations of sizable
elliptic flow and the constituent quark scaling behavior for the
multi-strange baryons suggest that substantial collective motion
has been developed prior to hadronization in the high-energy
nuclear collisions at RHIC.


We thank the RHIC Operations Group and RCF at BNL, and the NERSC
Center at LBNL for their support. This work was supported in part by
the HENP Divisions of the Office of Science of the U.S. DOE; the
U.S. NSF; the BMBF of Germany; IN2P3, RA, RPL, and EMN of France;
EPSRC of the United Kingdom; FAPESP of Brazil; the Russian Ministry
of Science and Technology; the Ministry of Education and the NNSFC
of China; IRP and GA of the Czech Republic, FOM of the Netherlands,
DAE, DST, and CSIR of the Government of India; Swiss NSF; the Polish
State Committee for Scientific Research; and the STAA of Slovakia.



\vfill\eject
\end{document}

%% file: sci-apr05.tex
\affiliation{Argonne National Laboratory, Argonne, Illinois 60439}
\affiliation{University of Bern, 3012 Bern, Switzerland}
\affiliation{University of Birmingham, Birmingham, United Kingdom}
\affiliation{Brookhaven National Laboratory, Upton, New York 11973}
\affiliation{California Institute of Technology, Pasadena, California 91125}
\affiliation{University of California, Berkeley, California 94720}
\affiliation{University of California, Davis, California 95616}
\affiliation{University of California, Los Angeles, California 90095}
\affiliation{Carnegie Mellon University, Pittsburgh, Pennsylvania 15213}
\affiliation{Creighton University, Omaha, Nebraska 68178}
\affiliation{Nuclear Physics Institute AS CR, 250 68 \v{R}e\v{z}/Prague, Czech Republic}
\affiliation{Laboratory for High Energy (JINR), Dubna, Russia}
\affiliation{Particle Physics Laboratory (JINR), Dubna, Russia}
\affiliation{University of Frankfurt, Frankfurt, Germany}
\affiliation{Institute of Physics, Bhubaneswar 751005, India}
\affiliation{Indian Institute of Technology, Mumbai, India}
\affiliation{Indiana University, Bloomington, Indiana 47408}
\affiliation{Institut de Recherches Subatomiques, Strasbourg, France}
\affiliation{University of Jammu, Jammu 180001, India}
\affiliation{Kent State University, Kent, Ohio 44242}
\affiliation{Lawrence Berkeley National Laboratory, Berkeley, California 94720}
\affiliation{Massachusetts Institute of Technology, Cambridge, MA 02139-4307}
\affiliation{Max-Planck-Institut f\"ur Physik, Munich, Germany}
\affiliation{Michigan State University, East Lansing, Michigan 48824}
\affiliation{Moscow Engineering Physics Institute, Moscow Russia}
\affiliation{City College of New York, New York City, New York 10031}
\affiliation{NIKHEF and Utrecht University, Amsterdam, The Netherlands}
\affiliation{Ohio State University, Columbus, Ohio 43210}
\affiliation{Panjab University, Chandigarh 160014, India}
\affiliation{Pennsylvania State University, University Park, Pennsylvania 16802}
\affiliation{Institute of High Energy Physics, Protvino, Russia}
\affiliation{Purdue University, West Lafayette, Indiana 47907}
\affiliation{University of Rajasthan, Jaipur 302004, India}
\affiliation{Rice University, Houston, Texas 77251}
\affiliation{Universidade de Sao Paulo, Sao Paulo, Brazil}
\affiliation{University of Science \& Technology of China, Anhui 230027, China}
\affiliation{Shanghai Institute of Applied Physics, Shanghai 201800, China}
\affiliation{SUBATECH, Nantes, France}
\affiliation{Texas A\&M University, College Station, Texas 77843}
\affiliation{University of Texas, Austin, Texas 78712}
\affiliation{Tsinghua University, Beijing 100084, China}
\affiliation{Valparaiso University, Valparaiso, Indiana 46383}
\affiliation{Variable Energy Cyclotron Centre, Kolkata 700064, India}
\affiliation{Warsaw University of Technology, Warsaw, Poland}
\affiliation{University of Washington, Seattle, Washington 98195}
\affiliation{Wayne State University, Detroit, Michigan 48201}
\affiliation{Institute of Particle Physics, CCNU (HZNU), Wuhan 430079, China}
\affiliation{Yale University, New Haven, Connecticut 06520}
\affiliation{University of Zagreb, Zagreb, HR-10002, Croatia}

\author{J.~Adams}\affiliation{University of Birmingham, Birmingham, United Kingdom}
\author{M.M.~Aggarwal}\affiliation{Panjab University, Chandigarh 160014, India}
\author{Z.~Ahammed}\affiliation{Variable Energy Cyclotron Centre, Kolkata 700064, India}
\author{J.~Amonett}\affiliation{Kent State University, Kent, Ohio 44242}
\author{B.D.~Anderson}\affiliation{Kent State University, Kent, Ohio 44242}
\author{D.~Arkhipkin}\affiliation{Particle Physics Laboratory (JINR), Dubna, Russia}
\author{G.S.~Averichev}\affiliation{Laboratory for High Energy (JINR), Dubna, Russia}
\author{S.K.~Badyal}\affiliation{University of Jammu, Jammu 180001, India}
\author{Y.~Bai}\affiliation{NIKHEF and Utrecht University, Amsterdam, The Netherlands}
\author{J.~Balewski}\affiliation{Indiana University, Bloomington, Indiana 47408}
% ...

\collaboration{STAR Collaboration}\noaffiliation